\documentclass[10pt,prd,aps,nofootinbib,floatfix,superscriptaddress,preprintnumbers]{revtex4}

\usepackage[skip=10pt plus1pt, indent=20pt]{parskip}
\usepackage{graphicx}
\usepackage{amsmath}
\usepackage{bbold}
\usepackage{orcidlink}
\graphicspath{{./figures/},{./plots/},{./plots/mixing_angle/},{./plots/spectrum/},{./plots/effective_mass/}}

\newcommand{\orcidauthorBENNETT}{0000-0002-1678-6701}
\newcommand{\orcidauthorFORZANO}{0000-0003-0985-8858}
\newcommand{\orcidauthorHONG}{0000-0002-3923-4184}
\newcommand{\orcidauthorHSIAO}{0000-0002-8522-5190}
\newcommand{\orcidauthorLEE}{0000-0002-4616-2422}
\newcommand{\orcidauthorLIN}{0000-0003-3743-0840}
\newcommand{\orcidauthorLUCINI}{0000-0001-8974-8266}
\newcommand{\orcidauthorPIAI}{0000-0002-2251-0111} 
\newcommand{\orcidauthorVADACCHINO}{0000-0002-5783-5602}
\newcommand{\orcidauthorZIERLER}{0000-0002-8670-4054}

\newcommand{\Tr}{{\rm Tr}}
\newcommand{\etalight}{{\eta^{\prime}_l}}
\newcommand{\etaheavy}{{\eta^{\prime}_h}}
\newcommand{\rev}[1]{{#1}}

\begin{document}

\title{On the mixing between flavor singlets in lattice gauge theories coupled to matter fields in multiple representations }

\author{Ed Bennett\,\orcidlink{\orcidauthorBENNETT}}
\email{E.J.Bennett@swansea.ac.uk}
\affiliation{Swansea Academy of Advanced Computing, Swansea University (Bay Campus), Fabian Way, SA1 8EN Swansea, Wales, United Kingdom}

\author{Niccolò Forzano\,\orcidlink{\orcidauthorFORZANO}}
\email{2227764@swansea.ac.uk}
\affiliation{Department of Physics, Faculty  of Science and Engineering, Swansea University, Singleton Park, SA2 8PP, Swansea, Wales, UK}

\author{Deog~Ki Hong\,\orcidlink{\orcidauthorHONG}}
\email{dkhong@pusan.ac.kr}
\affiliation{Department of Physics, Pusan National University, Busan 46241, Korea}
\affiliation{Extreme Physics Institute, Pusan National University, Busan 46241, Korea}

\author{Ho Hsiao\,\orcidlink{\orcidauthorHSIAO}}
\email{thepaulxiao.sc06@nycu.edu.tw}
\affiliation{Institute of Physics, National Yang Ming Chiao Tung University, 1001 Ta-Hsueh Road, Hsinchu 30010, Taiwan}

\author{Jong-Wan Lee\,\orcidlink{\orcidauthorLEE}}
\email{j.w.lee@ibs.re.kr}
\affiliation{ Particle Theory  and Cosmology Group, Center for Theoretical Physics of the Universe, Institute for Basic Science (IBS), Daejeon, 34126, Korea }

\author{C.-J. David Lin\,\orcidlink{\orcidauthorLIN}}
\email{dlin@nycu.edu.tw}
\affiliation{Institute of Physics, National Yang Ming Chiao Tung University, 1001 Ta-Hsueh Road, Hsinchu 30010, Taiwan}
\affiliation{Centre for Theoretical and Computational Physics, National Yang Ming Chiao Tung University, 1001 Ta-Hsueh Road, Hsinchu 30010, Taiwan }
\affiliation{Centre for High Energy Physics, Chung-Yuan Christian University, Chung-Li 32023, Taiwan}

\author{Biagio~Lucini\,\orcidlink{\orcidauthorLUCINI}}
\email{B.Lucini@Swansea.ac.uk}
\affiliation{Swansea Academy of Advanced Computing, Swansea University (Bay Campus), Fabian Way, SA1 8EN Swansea, Wales, United Kingdom}
\affiliation{Department of Mathematics, Faculty of Science and Engineering, Swansea University (Bay Campus), Fabian Way, SA1 8EN Swansea, Wales, United Kingdom}

\author{Maurizio Piai\,\orcidlink{\orcidauthorPIAI}}
\email{m.piai@swansea.ac.uk}
\affiliation{Department of Physics, Faculty  of Science and Engineering, Swansea University, Singleton Park, SA2 8PP, Swansea, Wales, UK}

\author{Davide Vadacchino\,\orcidlink{\orcidauthorVADACCHINO}}
\email{davide.vadacchino@plymouth.ac.uk}
\affiliation{Centre for Mathematical Sciences, University of Plymouth, Plymouth, PL4 8AA, United Kingdom}

\author{Fabian Zierler\,\orcidlink{\orcidauthorZIERLER}}
\email{fabian.zierler@swansea.ac.uk}
\affiliation{Department of Physics, Faculty  of Science and Engineering, Swansea University, Singleton Park, SA2 8PP, Swansea, Wales, UK}

\date{\today}

\begin{abstract}

We provide the first extensive,  numerical  study of the non-trivial problem of mixing between flavor-singlet composite states emerging in strongly coupled lattice field theories with matter field content consisting of fermions transforming in different representations of the gauge group. The theory of interest is the minimal candidate for a composite Higgs model that also accommodates a mechanism for top partial compositeness: the $Sp(4)$ gauge theory coupled to two (Dirac) fermions transforming as the fundamental and three as the two-index antisymmetric representation of the gauge group, respectively. We apply an admixture of APE and Wuppertal smearings, as well as the generalized eigenvalue problem approach, to two-point functions involving flavor-singlet  mesons, for ensembles having  time extent longer than the space extent.  We demonstrate that, in the region of lattice parameter space accessible to this study, both masses  and mixing angles can be measured effectively, despite the presence of (numerically noisy) contributions from disconnected diagrams.
\end{abstract}
\preprint{CTPU-PTC-24-12}
\preprint{PNUTP-24/A03}
\maketitle
\newpage
\tableofcontents

\section{Introduction}
\label{Sec:I}

A distinguishing feature of a broad class of models of new physics featuring composite dynamics is the emergence in their spectrum of several scalar singlets, associated with spontaneously broken, approximate, anomalous Abelian continuous symmetries~\cite{Belyaev:2016ftv}. In the context of composite Higgs models (CHMs)~\cite{Kaplan:1983fs,Georgi:1984af, Dugan:1984hq} that also implement top partial compositeness (TPC)~\cite{Kaplan:1991dc},\footnote{An incomplete catalog of such models, in which the Higgs fields of the Standard Model emerge as pseudo-Nambu-Goldstone bosons (PNGBs) associated with the spontaneous breaking of approximate global symmetries, and of studies of their phenomenology, includes for example Refs.~\cite{Barbieri:2007bh, Lodone:2008yy, Gripaios:2009pe, Mrazek:2011iu, Marzocca:2012zn,  Grojean:2013qca, Barnard:2013zea, Cacciapaglia:2014uja, Ferretti:2014qta, Arbey:2015exa, Cacciapaglia:2015eqa, vonGersdorff:2015fta, Feruglio:2016zvt, DeGrand:2016pgq, Fichet:2016xvs, Galloway:2016fuo, Agugliaro:2016clv, Bizot:2016zyu, Csaki:2017cep, Chala:2017sjk, Golterman:2017vdj, Serra:2017poj, Csaki:2017jby, Alanne:2017rrs, Alanne:2017ymh, Sannino:2017utc, Alanne:2018wtp, Bizot:2018tds, Cai:2018tet, Agugliaro:2018vsu, Cacciapaglia:2018avr, Gertov:2019yqo, Ayyar:2019exp, Cacciapaglia:2019ixa, BuarqueFranzosi:2019eee, Cacciapaglia:2019dsq, Cacciapaglia:2020vyf, Dong:2020eqy, Cacciapaglia:2021uqh, Ferretti:2022mpy, Banerjee:2022izw, Cai:2022zqu, Cacciapaglia:2024wdn}. Implementations of similar ideas within  gauge-gravity dualities have been presented for example in Refs.~\cite{Contino:2003ve, Agashe:2004rs, Agashe:2005dk, Agashe:2006at, Contino:2006qr, Falkowski:2008fz, Contino:2010rs, Contino:2011np} and, more recently, in Refs.~\cite{Erdmenger:2020lvq,Erdmenger:2020flu,Elander:2020nyd,Elander:2021bmt, Erdmenger:2023hkl,Elander:2023aow,Erdmenger:2024dxf}, in the bottom-up approach, and Ref.~\cite{Elander:2021kxk} in the top-down approach to holography.}
the multiplicity of such flavor-singlet states arises because the short-distance origin of the dynamics involves two separate matter sectors coupled to the same gauge theory.\footnote{This requirement may not apply in $SU(3)$ theories, for which the ordinary baryons may provide the origin of the TPC fields~\cite{Vecchi:2015fma}--- see also the constructions in Refs.~\cite{Appelquist:2020bqj, Appelquist:2022qgl}, which make use of ideas from Refs.~\cite{Ma:2015gra,BuarqueFranzosi:2018eaj}} Examples can be found in the reviews~\cite{Panico:2015jxa,Witzel:2019jbe,Cacciapaglia:2020kgq}, and the tables in Refs.~\cite{Ferretti:2013kya,Ferretti:2016upr,Cacciapaglia:2019bqz}. Closely related classes of theories are also known to admit an application as new models of dark matter with strongly coupled origin, for example along the lines of Refs.~\cite{Hochberg:2014dra,Hochberg:2014kqa,Hochberg:2015vrg,Hansen:2015yaa, Bernal:2017mqb,Berlin:2018tvf, Bernal:2019uqr,Beylin:2019gtw,Tsai:2020vpi,Kondo:2022lgg,Bernal:2015xba,Pomper:2024otb,Appelquist:2024koa}.

The existence of mixing terms in the effective field theory (EFT) description of such flavor-singlets is ensured by the chiral anomaly, which breaks the symmetry even when no other explicit symmetry-breaking terms are present. The phenomenology associated with such scalar singlets is determined by coefficients in the EFT treatment that have dynamical origin---related to that of axion-like particles (ALPs)~\cite{Brivio:2017ije,Bellazzini:2017neg,Bauer:2017ris}. This is the case in the dark matter context, as in the collider phenomenology one~\cite{Belyaev:2016ftv,Cai:2015bss,Belyaev:2015hgo,DeGrand:2016pgq, Cacciapaglia:2017iws,Cacciapaglia:2019bqz}. Hence, gaining non-perturbative information about them is essential in order to plan and perform an effective program of experimental searches for new particles in both visible and dark sectors. 

In recent years, extensive numerical investigations have been developed in the context of extensions to the Standard Model, based on symplectic gauge groups $SU(2)=Sp(2)$~\cite{Hietanen:2014xca,Detmold:2014kba,Drach:2015epq,Arthur:2016dir,Arthur:2016ozw, Pica:2016zst,Lee:2017uvl,Drach:2017jsh,Drach:2017btk,Drach:2020wux,Drach:2021uhl,Bowes:2023ihh} and $Sp(4)$~\cite{Bennett:2017kga,Lee:2018ztv,Bennett:2019jzz,Bennett:2019cxd, Lucini:2021xke,Bennett:2021mbw,Bennett:2022yfa,Lee:2022xbp,Bennett:2023wjw,Bennett:2023rsl,Bennett:2023gbe,Mason:2023ixv,Zierler:2021cfa,Kulkarni:2022bvh,Bennett:2023mhh,Bennett:2023qwx,Hsiao:2023nyn}. Among the gauge theories with fermions in two distinct representations, lattice studies have been performed in $SU(2)$~\cite{Bergner:2020mwl,Bergner:2021ivi}, as well as in $Sp(4)$ gauge theories~\cite{Bennett:2022yfa,Hsiao:2022kxf,Bennett:2023mhh,Hsiao:2023nyn}. In a parallel development, studies of theories with  fermions transforming in multiple representation of $SU(4)$ have also appeared~\cite{Ayyar:2017qdf,Ayyar:2018zuk,Ayyar:2018ppa,Ayyar:2018glg,Cossu:2019hse,Shamir:2021frg,Lupo:2021nzv,DelDebbio:2022qgu,Hasenfratz:2023sqa}. These lattice studies, motivated by new physics considerations, focus predominantly on the study of flavored mesons. \footnote{The literature on the $SU(3)$ theory with $N_{\rm f}=8$ Dirac fermions in the fundamental representation ~\cite{LatKMI:2014xoh, Appelquist:2016viq,LatKMI:2016xxi, Gasbarro:2017fmi,LatticeStrongDynamics:2018hun,LSD:2023uzj,LatticeStrongDynamics:2023bqp,Ingoldby:2023mtf} or $n_{\rm s}=2$ sextets~\cite{Fodor:2012ty,	Fodor:2015vwa, Fodor:2016pls,Fodor:2017nlp,Fodor:2019vmw,Fodor:2020niv}, stands out as it presents extensive studies of the flavor singlet states. But these studies focus on  the phenomenology of the dilaton, the Goldstone boson associated with scale invariance, which has different quantum numbers from the PNGBs associated with Abelian internal symmetries.}
This lattice subfield is only beginning to enter its high precision phase, after the necessary exploratory period. Calculating correlation functions involving singlet mesons requires specific technology, developed to handle effectively the contributions to observables coming from disconnected diagrams, which potentially reduce the signal-to-noise ratio. Only few studies in the context of BSM physics outside of $SU(3)$ theories exist ~\cite{ Drach:2021uhl, Arthur:2016ozw, Bennett:2023rsl}. 
To the best of our knowledge, this paper is the first one to present a systematic lattice study of the effects of mixing between different flavor singlets in models of BSM physics. Our aim is to demonstrate the feasibility of such an endeavor.

In this publication, we present the first results of the calculation of the mass spectrum of spin-0, flavor singlet states with negative parity, in the presence of two distinct Abelian PNGBs associated with two $U(1)$ factors. One of them is expected to be anomalously broken, thus giving an extra contribution to the mass of one PNGB \cite{Belyaev:2016ftv}. 
The lattice ensembles we study are obtained in the CHM candidate of Ref.~\cite{Barnard:2013zea}: the $Sp(4)$ gauge theory  coupled to $N_{\rm f}=2$ (Dirac) fermions transforming according to the fundamental, as well as $N_{\rm as}=3$ fermions transforming as the 2-index antisymmetric representation of the group. We use the ensembles described in detail in Ref.~\cite{Bennett:2024cqv}, originally produced to study the spectral density of  flavored meson correlators, within the Hansen-Lupo-Tantalo method~\cite{Hansen:2019idp,DelDebbio:2022qgu}. Building on the results discussed in the Appendix of Ref.~\cite{Bennett:2023rsl}, we introduce Wuppertal~\cite{Gusken:1989ad, Gusken:1989qx,Roberts:2012tp,Alexandrou:1990dq} and APE smearing~\cite{APE:1987ehd,Falcioni:1984ei}. We then implement a variation of the Generalized Eigenvalue Problem (GEVP) to take into account  mixing effects between states that appear in two-point functions  involving the two distinct meson singlets in the theory.

The paper is organized as follows.
We very briefly introduce the continuum and lattice theories of interest in Sects.~\ref{Sec:L1} and~\ref{Sec:L2}, respectively.  We provide the minimal amount of detail to make the narrative self-contained, and refer the Reader to Ref.~\cite{Bennett:2024cqv} for technical details and in-depth discussions.
Nevertheless, we extensively describe the properties of the flavor-singlet meson  states of interest, with 
particular reference to decay constants and mixing angles,  as well as  our analysis techniques,
in Sect.~\ref{Sec:M}. Our results are then reported and critically discussed in Sect.~\ref{Sec:R}.
A final summary in Sect.~\ref{Sec:S} contains also a brief outlook on future avenues for research.

\section{Elements of field theory}
\label{Sec:L1}

We study the $Sp(4)$ gauge theory coupled to two Dirac fermions transforming in the fundamental representation and three Dirac fermions in the antisymmetric representation of the gauge group. The Lagrangian density, in Minkowski space, is given by
\begin{align}
    \mathcal L = -\frac{1}{2} \Tr G_{\mu\nu} G^{\mu\nu} +  \sum_{I=1}^{2} \bar Q^I \left( i\gamma^\mu D_\mu - m^{\rm f} \right) Q^I+ \sum_{K=1}^{3} \bar \Psi^k \left( i\gamma^\mu D_\mu - m^{\rm as} \right) \Psi^k,
\end{align}
where $Q^I$, with $I=1,\,2$, are the fundamental Dirac fermions and $\Psi^k$, with $k=1,\,2,\,3$, are the antisymmetric Dirac fermions with their respective degenerate masses $m^{\rm f}$ and $m^{\rm as}$. The field-strength tensor for the $Sp(4)$ theory is denoted as $G_{\mu\nu}\equiv G_{\mu\nu}^aT^a$, with the Hermitian generators normalized so that $\Tr T^aT^b=\frac{1}{2} \delta^{ab}$, for $a,\,b=1,\,\cdots,\, 10$.

In the limit of vanishing fermion masses, the Lagrangian density is invariant under the enhanced global internal symmetry with group 
$U(1) \times U(1) \times SU(6) \times SU(4)$. This symmetry of the Lagrangian is not reflected in low-energy physics; it is broken spontaneously by the fermion condensates. The breaking pattern is governed by the realness of the antisymmetric representation and the pseudo-realness of the fundamental representation; $SU(6)$ breaks to its $SO(6)$ subgroup, while $SU(4)$ breaks to $Sp(4)$~\cite{Kosower:1984aw}. At the same time, the diagonal and degenerate mass terms induce explicit symmetry breaking, according to the same, aligned  symmetry breaking pattern. For small fermion masses, the spontaneous breaking leads to 20 PNGBs in the antisymmetric sector and 5 PNGBs in the fundamental sector~\cite{Kogut:2000ek}. In the CHM context, one can choose 
appropriate embeddings for the standard-model gauge group so that part of the global symmetry is (weakly) gauged.
 For example, in this way the five PNGBs of the fundamental sector can be identified with the Higgs doublet and one additional singlet~\cite{Barnard:2013zea}. In this paper we consider the $Sp(4)$ gauge theory in complete isolation, 
 ignoring the effects due to coupling to other external sectors.

In theories with  only one species of fermions, the global U(1) is anomalous, and no Goldstone bosons, related to its breaking, appear. In the large-$N$ limit, the associated pseudoscalar flavor-singlet, $\eta'$, becomes the would-be Goldstone mode, as the effects of the axial anomaly are
suppressed in this limit~\cite{Witten:1978bc,Witten:1979vv,Veneziano:1979ec}. 
In the presence of fermions in two distinct representations, the axial anomaly can only break one (linear combination) of the two global $U(1)$ symmetry factors. Hence, an additional PNGB is expected to appear at small fermion masses, in the pseudoscalar flavor-singlet sector,  the lightest mass eigenstate in this channel. The other pseudoscalar flavor-singlet should acquire a larger mass through the axial anomaly and show up as an excited state in the 
same channel~\cite{Belyaev:2015hgo}. 

In this paper, we perform the first measurement of flavor-singlet ground and excited states in a theory with multiple fermion representations, which we call $\etalight$ and $\etaheavy$\footnote{Note, that this corresponds to the states $a$ and $\eta'$ in the notation of Ref.~\cite{Belyaev:2016ftv}}. The theory of interest has $Sp(4)$ gauge group, and explicit mass terms for the fermions, which are not small. Hence, we expect both the aforementioned flavor singlets to be comparatively heavy, as the mass terms contribute together with the anomaly to the explicit symmetry breaking.
One can draw an analogy between this regime and 
the $\eta-\eta'$ system in the theory of strong nuclear interactions, QCD \footnote{Note, that this analogy has its limits. In QCD, the mixing between the $\eta$ and the $\eta'$ is introduced by an explicit breaking of the global flavor symmetry. For the mixed representations system, this is not the case, and both states are pseudoscalar singlets even without additional symmetry breaking.}. In this case, the $\eta$ meson is a PNGB accommodated within the approximate 
$SU(3)$ flavor symmetry, while the $\eta'$ is the would-be-PNGB associated with the anomalously broken global $U(1)$. Within QCD, both the mass spectrum and the mixing between those states have been studied in detail \cite{Christ:2010dd,Dudek:2011tt,Ottnad:2012fv,Michael:2013gka,Bali:2014pva,Ottnad:2015hva,Ottnad:2017bjt,Bali:2021qem,CSSMQCDSFUKQCD:2021rvs,Bali:2021vsa,Simeth:2022fuq}.

\section{Numerical Strategy}
\label{Sec:L2}

Our numerical analysis is based on ensembles generated using the Wilson plaquette action for the gauge sector and standard Wilson fermions for both the fundamental and antisymmetric fermions~\cite{Wilson:1974sk}. We consider hypercubic lattices with a volume $L^3 \times T = a^4 (N_s^3 \times N_t)$, where $a$ is the lattice spacing. The ensembles have been generated using the Grid software library~\cite{Boyle:2015tjk,Boyle:2016lbp,Yamaguchi:2022feu} which has been extended to $Sp(2N)$ gauge theories \cite{Bennett:2023gbe}. We study five ensembles, at three different values of the bare fundamental fermion mass, $am_0^{\rm f}=-0.7,\,-0.71,\,0.72$, while we keep the antisymmetric fermion mass fixed, $am_0^{\rm as}=-1.01$. We consider only one value of the inverse gauge coupling, $\beta = 8/g^2 = 6.5$. The measurements of the mesonic correlation function are performed using the HiRep code~\cite{DelDebbio:2008zf,HiRepSUN,HiRepSpN}. The configurations have been converted to the HiRep binary format using the GLU library~\cite{GLU}. We set the overall scale using the Wilson flow~\cite{Luscher:2010iy}, the lattice implementation of the gradient flow~\cite{Luscher:2011bx,Luscher:2013vga}, by calculating the gradient flow quantity, $w_0$~\cite{BMW:2012hcm}, in units of the lattice spacing, $a$. For further information on the generation of these ensembles we refer to Ref.~\cite{Bennett:2024cqv}. Detailed characterization of the five ensembles is summarized in Tab.~\ref{tab:ensembles}.

\begin{table}
    \caption{Ensembles studied in this paper. For each of the five ensembles, we list the value of the inverse lattice coupling, $\beta$, the  masses of the two fermion species, $am_0^{\rm as}$ and $am_0^{\rm f}$, the extent of the lattice in time, $N_t$, and space directions, $N_s$,  the number of thermalization steps, $N_{\rm therm}$, discarded from the analysis, the number of complete sweeps, $n_{\rm skip}$, discarded between configurations retained in the analysis, the number of configurations constituting the ensemble and used in the analysis,   $N_{\rm conf}$, the average plaquette in the ensemble, $\langle P \rangle$, the value of the Wilson flow scale, $w_0 / a$, the topological autocorrelation time in configuration units, $\tau_{\rm int}^Q$, and the average topological charge in the ensemble,  $\bar{Q}$. Details, explanations and discussions can be found in Ref.~\cite{Bennett:2024cqv}.
    \label{tab:ensembles}}
    \begin{tabular}{|c|c|c|c|c|c|c|c|c|c|c|c|c|}
    \hline \hline
    Label & $\beta$ & $am_0^{\rm as}$ & $am_0^{\rm f}$ & $N_t$ & $N_s$ & $N_{\rm therm}$ & $n_{\rm skip}$ & $N_{\rm conf}$ & $\langle P \rangle$ & $w_0 / a$ & $\tau_{\rm int}^Q$ &  $\bar{Q}$ \\ 
    \hline
        M1 & 6.5 & -1.01 & -0.71 & 48 & 20 & 3006 & 14 & 479 & 0.585172(16) & 2.5200(50) & 6.9(2.4) & 0.38(12)\\ 
        M2 & 6.5 & -1.01 & -0.71 & 64 & 20 & 1000 & 28 & 698 & 0.585172(12) & 2.5300(40) & 7.1(2.1) & 0.58(14)\\ 
        M3 & 6.5 & -1.01 & -0.71 & 96 & 20 & 4000 & 26 & 436 & 0.585156(13) & 2.5170(40) & 6.4(3.3) & -0.60(19)\\ 
        M4 & 6.5 & -1.01 & -0.70 & 64 & 20 & 1000 & 20 & 709 & 0.584228(12) & 2.3557(31) & 10.6(4.8) & -0.31(19)\\ 
        M5 & 6.5 & -1.01 & -0.72 & 64 & 32 & 3020 & 20 & 295 & 0.5860810(93) & 2.6927(31) & 12.9(8.2) & 0.80(33)\\ 
    \hline \hline
    \end{tabular}
\end{table}

We determine the ground state energy as well as the energy of the first excited state for the system of interest by using a variational analysis~\cite{Michael:1985ne, Luscher:1990ck}. To this purpose, we measure the (zero-momentum) correlation matrix  for a set of interpolating operators, $\left\{ O_i \right\}$, which can be expanded in the Hamiltonian eigenstates as
\begin{align}\label{eq:correlation_matrix}
    C_{ij}(t-t') = \langle \bar O_i(t) O_j(t') \rangle = \sum_n \frac{1}{2E_n} \langle 0 \vert \bar O_i \vert n \rangle \langle n \vert  O_j \vert 0 \rangle e^{-E_n (t-t')} .
\end{align}
The eigenvalues of $C_{ij}$ are obtained by solving the generalized eigenvalue problem (GEVP) with eigenvalues $\lambda_n(t,t_0)$ and eigenvectors $v_n(t,t_0)$ given by
\begin{align}\label{eq:GEVP}
    C(t) v_n(t,t_0) = \lambda_n(t,t_0) C(t_0) v_n(t,t_0).
\end{align}
Assuming that the states created by each of the operators in the variational basis has a sufficient overlap with the first $M$ eigenstates in the selected channel, the leading behavior of the $m^{\rm th}$ eigenvalue at large $t$ for fixed but sufficiently large $t_0$ can be written as \cite{Blossier:2009kd}
\begin{align}\label{eq:eigenvalue_asymptotics}
    \lambda_m(t \to \infty,t_0) =  e^{-E_m (t-t_0)} + \mathcal{O} \left( e^{-(E_{m+1} - E_m)t} \right).
\end{align}
In summary, the solution of the above GEVP enables us to find the operator that produces states with the maximal overlap with the ground state and the first excited state, and thus to access their energies.

The two meson operators of interest in the pseudoscalar singlet channel are coupled, respectively, to $N_{\rm f}$ fundamental fermions and $N_{\rm as}$ antisymmetric fermions, and they are given by the following\footnote{We note, that in principle a contribution from the $J^P=0^-$ glueball state is also allowed. However, an investigation of $\eta'$-glueball mixing in two-flavor QCD showed no sizeable contributions from the glueball state~\cite{Jiang:2022ffl}. Studies of the quenched $Sp(4)$ theory suggest that indeed the $0^-$ glueballs is very heavy with respect to the scale of the vector mesons---see Fig.~13 of Ref.~\cite{ Bennett:2023mhh}, for example. }:
\begin{align} \label{eq:ps_operator1}
    O_{\eta^{\rm as}}(x) &= \frac{1}{\sqrt{N_{\rm as}}} \sum_{k=1}^{N_{\rm as}} \bar \Psi^k(x) \gamma_5 \Psi^k(x)\,, \\
     \label{eq:ps_operator2}
    O_{\eta^{\rm f}}(x)  &= \frac{1}{\sqrt{N_{\rm f}}} \sum_{I=1}^{N_{\rm f}} \bar  Q^I(x)   \gamma_5  Q^I(x)\,.
\end{align}
Performing the required Wick contractions for the  correlation matrix resulting from Eq.~\eqref{eq:correlation_matrix}, for operators at lattice sites $x$ and $y$, we find  the diagrammatic expression
\begin{align}
    &C_{11}(x,y) \equiv \langle \bar O_{\eta^{\rm as}}(x) O_{\eta^{\rm as}}(y) \rangle = 
    - \vcenter{\hbox{\includegraphics[scale=0.5,page=31]{contractions.pdf}}}
     + N_{\rm as}~\vcenter{\hbox{\includegraphics[scale=0.5,page=33]{contractions.pdf}}}, \label{eq:wick1} \\
    &C_{22}(x,y) \equiv \langle \bar O_{\eta^{\rm  f}}(x) O_{\eta^{\rm  f}}(y) \rangle =
     - \vcenter{\hbox{\includegraphics[scale=0.5,page=30]{contractions.pdf}}}
      + N_{\rm f}~\vcenter{\hbox{\includegraphics[scale=0.5,page=34]{contractions.pdf}}}, \label{eq:wick2}\\ 
    &C_{12}(x,y) = C_{21}(x,y) \equiv \langle \bar O_{\eta^{\rm f}}(x) O_{\eta^{\rm as}}(y) \rangle =  + \sqrt{N_{\rm as} N_{\rm f} } \vcenter{\hbox{\includegraphics[scale=0.5,page=32]{contractions.pdf}}}, \label{eq:wick3}
\end{align}
where dashed lines denote the contraction of two antisymmetric fermion fields and solid lines denote the contraction of two fundamental fermion fields. Unlike the singlet-octet basis in QCD for the $\eta$ and $\eta'$ mesons, the cross-correlator does not vanish in the limit of vanishing fermion masses, hence we expect sizeable mixing effects for moderate and light fermion masses.

The disconnected diagrams in Eqs.~\eqref{eq:wick1}-\eqref{eq:wick3} are challenging for lattice calculations, as they introduce  a smaller signal-to-noise ratio than the connected diagrams---that are also present in non-singlet mesons. Our objective is to improve the  signal by enlarging the variational basis through smearing techniques. The general principle at work is that one expects the adoption of larger variational basis in the GEVP analysis to suppress the effects of excited state contamination at smaller Euclidean times, where the signal-to-noise ratio is substantially better.

We implement Wuppertal smearing of the fermionic operators~\cite{Gusken:1989ad, Gusken:1989qx,Roberts:2012tp,Alexandrou:1990dq}, in conjunction with APE smearing for the gauge fields~\cite{APE:1987ehd,Falcioni:1984ei}. We follow the approach used in Ref.~\cite{Bennett:2023mhh} for the connected diagrams, and apply the smearing function to point sources. For the disconnected diagrams, we use spin-diluted stochastic sources~\cite{Foley:2005ac}, with $Z_2 \times Z_2$ noise~\cite{Dong:1993pk}, and perform the measurements on $n_{\rm src}=64$ stochastic samples. As pointed out in Ref.~\cite{Bali:2021qem}, it is possible to measure the disconnected loops at several smearing levels using only one inversion of the Dirac operator. Doing so comes at the cost of potentially introducing a bias in the construction of the full correlation function, following the procedure deployed in Ref.~\cite{Arthur:2016ozw}. The construction of a completely unbiased estimator would require to perform a separate inversion for every smearing level, but this would go beyond the purposes of this paper.
We apply APE smearing for every measurement. We choose the smearing parameter as $\alpha_{\rm APE}=0.4$, with $N_{\rm APE}=50$ smearing steps. 

The variational basis is composed of the operators defined in Eq.~\eqref{eq:ps_operator1} and Eq.~\eqref{eq:ps_operator2} together with their Wuppertal-smeared versions, with $N^{\rm smear}=10,\,20,\,\dots,\,80$ smearing steps, and  $\epsilon_{\rm f}=0.2$ for the fundamental fermions, or $\epsilon_{\rm as}=0.12$ for the antisymmetric fermions (we follow the notation of Ref.~\cite{Bennett:2023mhh}, to which we refer the reader for technical details). In total, we have a variational basis of 18 operators. 

For a finite sample, the correlation functions of pseudoscalar flavor-singlet mesons (or any other quantity that has the same quantum numbers as the topological charge density) can acquire an additional, constant contribution~\cite{Aoki:2007ka}. In order to remove this constant we consider the central difference approximation to the derivative of the correlation matrix, as proposed in Ref.~\cite{Umeda:2007hy} and use of the following redefinition:
\begin{align}
    C_{ij}(t) \to \tilde C_{ij}(t) = \frac{C_{ij}(t-1) - C_{ij}(t+1)}{2}\,.
\end{align}
Doing so changes the periodicity of the correlation matrix with respect to the lattice midpoint, $t=T/2$, from periodic to anti-periodic~\cite{Bennett:2023rsl}. 

After performing the GEVP analysis, we fit the eigenvalues to an exponentially decaying function according to Eq.~\eqref{eq:eigenvalue_asymptotics}. We first visually examine the effective mass, $m_{\rm eff}(t)$, defined implicitly as
\begin{align}\label{eq:effective_mass}
    \frac{\lambda(t-1)}{\lambda(t)} = \frac{e^{- m_{\rm eff}(t) \cdot (T-t+1)} \pm e^{- m_{\rm eff}(t) \cdot (t-1)}}{e^{- m_{\rm eff}(t) \cdot (T-t)} \pm e^{- m_{\rm eff}(t) \cdot t}}\,.
\end{align}
We solve for the effective mass numerically using a root finding algorithm. It exhibits a plateau when the eigenvalue is dominated by its leading exponential term at large Euclidean time. The sign in Eq.~\eqref{eq:effective_mass} is chosen to be positive for symmetric correlation functions, and negative for antisymmetric ones. We then perform a fit to the eigenvalue using constrained curve 
fitting~\cite{Lepage:2001ym}, utilizing the corrfitter package~\cite{peter_lepage_2021_5733391}.

Finally, in order to gauge the physical meaning of the results of our analysis,
we use the same ensembles and processes also to perform the measurement of the mass of the
 lightest flavored mesons,
in the pseudoscalar and vector channel, for mesons constructed with either species of fermions.
We denote as ${\rm PS}~({\rm ps})$ flavored pseudoscalar mesons made of vfermions transforming in the fundamental, ${\rm f}$, (antisymmetric, ${\rm as}$) representation, and as  ${\rm V}~({\rm v})$ flavored vector mesons with the same composition. We refer the reader  to Ref.~\cite{Bennett:2024cqv} for more details about the operators, and the spectrum of flavored mesons, that are not central to the results and discussions presented in this paper.

\subsection{Decay constants and mixing angles}
\label{Sec:M}

It is interesting to extract  the mixing angle between the lightest states sourced by the operators in Eqs.~\eqref{eq:ps_operator1}  and~\eqref{eq:ps_operator2}. For the purposes of this paper, we assume that the state mixing is given by the mixing of the decay constants. 
We call $\etalight$ and $\etaheavy$, respectively, the lightest and next-to-lightest states identified in the GEVP analysis.
We parameterize the non-renormalized matrix elements of axial-vector currents with the pseudoscalar singlets as follows:
\begin{align} \label{eq:matrix_elements}
    \begin{pmatrix}
         F_{\rm f}^{\etalight} p_\mu^{\etalight} & F_{\rm as}^{\etalight} p_\mu^{\etalight}  \\
         F_{\rm f}^{\etaheavy} p_\mu^{\etaheavy} & F_{\rm as}^{\etaheavy} p_\mu^{\etaheavy}
    \end{pmatrix}
    \equiv
    \begin{pmatrix}
        \langle 0  | \frac{1}{\sqrt{N_{\rm f}}} \sum_{i=1}^{N_{\rm f}} \bar  Q_i(x) \gamma_\mu  \gamma_5  Q_i(x) | \etalight \rangle  & \langle 0  | \frac{1}{\sqrt{N_{\rm as}}} \sum_{i=1}^{N_{\rm as}} \bar \Psi_i(x) \gamma_\mu \gamma_5 \Psi_i(x) | \etalight \rangle   \\
        \langle 0  |  \frac{1}{\sqrt{N_{\rm f}}} \sum_{i=1}^{N_{\rm f}} \bar  Q_i(x) \gamma_\mu \gamma_5  Q_i(x) | \etaheavy \rangle  & \langle 0  |  \frac{1}{\sqrt{N_{\rm as}}} \sum_{i=1}^{N_{\rm as}} \bar \Psi_i(x) \gamma_\mu \gamma_5 \Psi_i(x) | \etaheavy \rangle
    \end{pmatrix}\,.
\end{align}
In this general relation, we define the relevant decay constants computed with an operator ($op$) and a state ($s$) as  $F^s_{op}$. By setting $\mu=0$ we obtain the familiar relation
\begin{align}
    \begin{pmatrix}
         F_{\rm f}^{\etalight} p_0^{\etalight} & F_{\rm as}^{\etalight} p_0^{\etalight}  \\
         F_{\rm f}^{\etaheavy} p_0^{\etaheavy} & F_{\rm as}^{\etaheavy} p_0^{\etaheavy}
    \end{pmatrix} = 
    \begin{pmatrix}
         F_{\rm f}^{\etalight} m_{\etalight} & F_{\rm as}^{\etalight} m_{\etalight}  \\
         F_{\rm f}^{\etaheavy} m_{\etaheavy} & F_{\rm as}^{\etaheavy} m_{\etaheavy}
    \end{pmatrix}.
\end{align}

These non-renormalized local matrix elements can be obtained from the eigenvectors of the GEVP analysis, for a variational basis without Wuppertal smearing. This matrix is parameterized as follows, in terms 
of two mixing angles, and two decay constants~\cite{Kaiser:1998ds}:
\begin{align}
    \begin{pmatrix} \label{eq:decay_constant_parametrisation}
         F_{\rm f}^{\etalight} & F_{\rm as}^{\etalight} \\
         F_{\rm f}^{\etaheavy} & F_{\rm as}^{\etaheavy}
    \end{pmatrix} \equiv
    \begin{pmatrix}
           F_{\etalight} \cos{\phi_\etalight} & F_{\etalight} \sin{\phi_\etalight} \\
         - F_{\etaheavy} \sin{\phi_\etaheavy} & F_{\etaheavy} \cos{\phi_\etaheavy}
    \end{pmatrix}\,,
\end{align}
in analogy to the $\eta-\eta'$ mixing system  in the singlet-octet basis of QCD. 
 
For this work, we choose a variational basis of interpolating operators and the matrix elements of the currents according to Eqs.~\eqref{eq:ps_operator1}, \eqref{eq:ps_operator2}, and~\eqref{eq:matrix_elements}. The difference in mass between mesons made of antisymmetric and fundamental fermions is expected to be large for the ensembles studied in this paper, on the basis of existing numerical results for other parts of the spectrum~\cite{Bennett:2022yfa}. In QCD, the two  mixing angles are approximately equal, $\phi_\etalight \approx \phi_{\etaheavy}$, at the physical point~\cite{Feldmann:1998vh} and Eq.~\eqref{eq:decay_constant_parametrisation} is effectively described by a single mixing angle, $\phi\equiv \phi_\etalight = \phi_{\etaheavy}$. We test this assumption for the present theory in Sec.~\ref{Sec:results_angle}.

A variation of this approach requires to use  the pseudoscalar matrix elements. This alternative approach has also been successfully applied to studying  $\eta-\eta'$ mixing in QCD~\cite{Christ:2010dd,Dudek:2011tt,Ottnad:2012fv,Michael:2013gka}. Again, the matrix elements are generically parameterized in terms of two mixing angles:
\begin{align} \label{eq:PS_matrix_elements}
    \begin{pmatrix}
        \langle 0  | O_{\eta^{\rm f }} | \etalight \rangle & \langle 0  | O_{\eta^{\rm as}} | \etalight \rangle    \\
        \langle 0  | O_{\eta^{\rm f}}  | \etaheavy \rangle & \langle 0  | O_{\eta^{\rm as}} | \etaheavy \rangle 
    \end{pmatrix}
    =
    \begin{pmatrix}
         A_{\rm f}^{\etalight} & A_{\rm as}^{\etalight}      \\
         A_{\rm f}^{\etaheavy} & A_{\rm as}^{\etaheavy}
    \end{pmatrix}
    \equiv
        \begin{pmatrix}
           A_{\etalight} \cos{\phi_\etalight} &  A_{\etalight} \sin{\phi_\etalight} \\
         - A_{\etaheavy} \sin{\phi_\etaheavy}  & A_{\etaheavy} \cos{\phi_\etaheavy}
    \end{pmatrix}\,.
\end{align}
We can extract the matrix elements of interest from the eigenvectors that diagonalize the $2\times2$  matrix of correlation functions defined by the local operators (i.e., without smearing). Under the assumption that only two states contribute to the $2\times 2$ correlation matrix, the eigenvectors, $v_n(t,t_0)$, obtained from the GEVP analysis 
should be proportional to time-independent vectors, $u_n$~\cite{Blossier:2009kd}. We use this to determine the matrix elements $F^s_{op} m_s$ and $A^s_{op}$, by fitting the \textit{effective mixing angle} $\phi$ defined as
\begin{align}\label{eq:effective_mixing_angle}
    - \left( \tan \phi \right)^2 \equiv \frac{A_{\rm as}^{\etalight} A_{\rm f}^{\etaheavy} }{A_{\rm as}^{\etaheavy} A_{\rm f}^{\etalight} }\,,
\end{align}
to a constant value,  for a suitable interval $t \in \left[ t_{\rm min}, t_{\rm max}\right]$, with $t_{\rm min} > t_0$. We \rev{further determine $\phi_\etalight$ and $\phi_\etaheavy$ independently to test the deviations from $\phi$ in a two-angle parametrization}. 

\section{Results}
\label{Sec:R}

\begin{figure}[ht!]
    \centering
    \includegraphics[width=.45\textwidth]{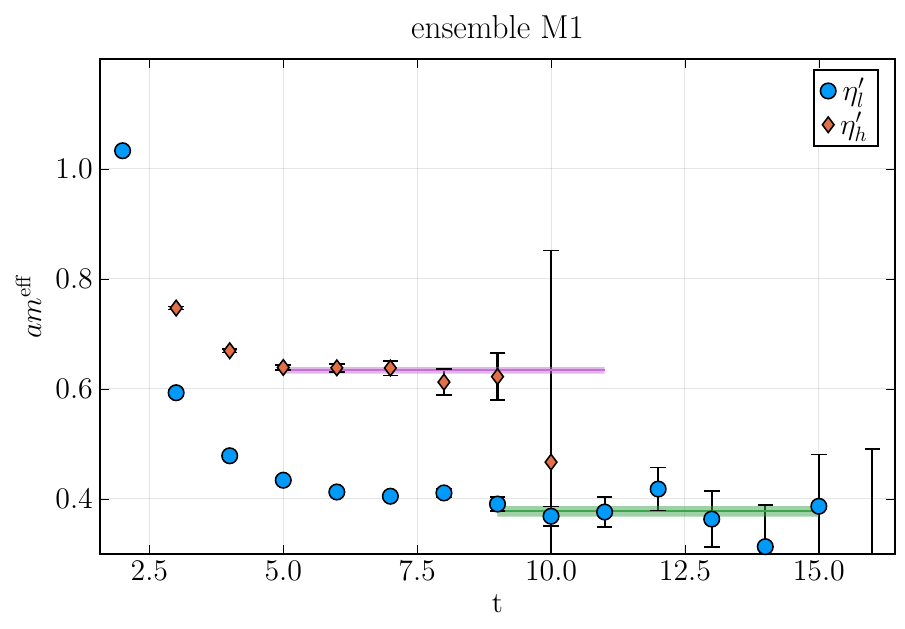}
    \includegraphics[width=.45\textwidth]{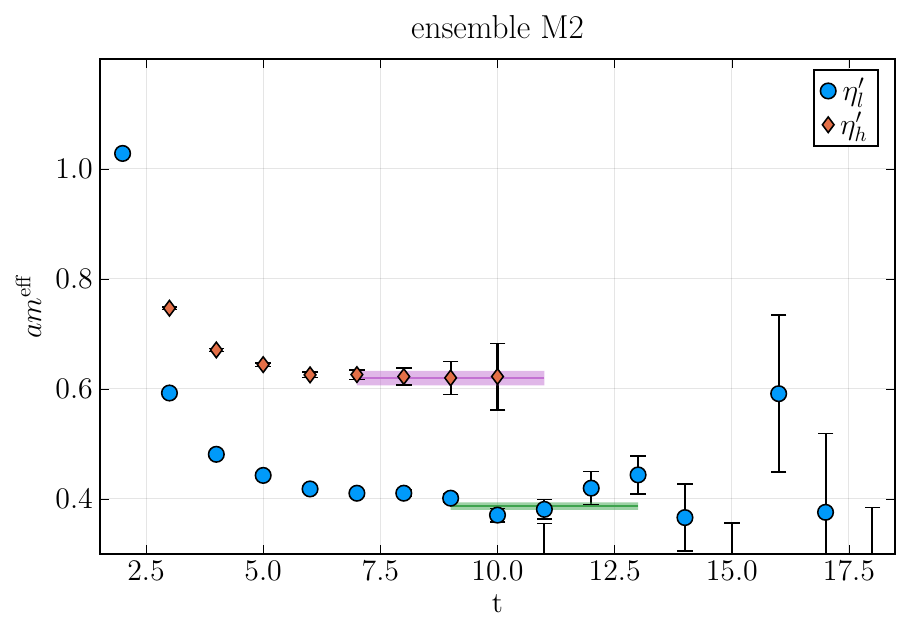}
    \includegraphics[width=.45\textwidth]{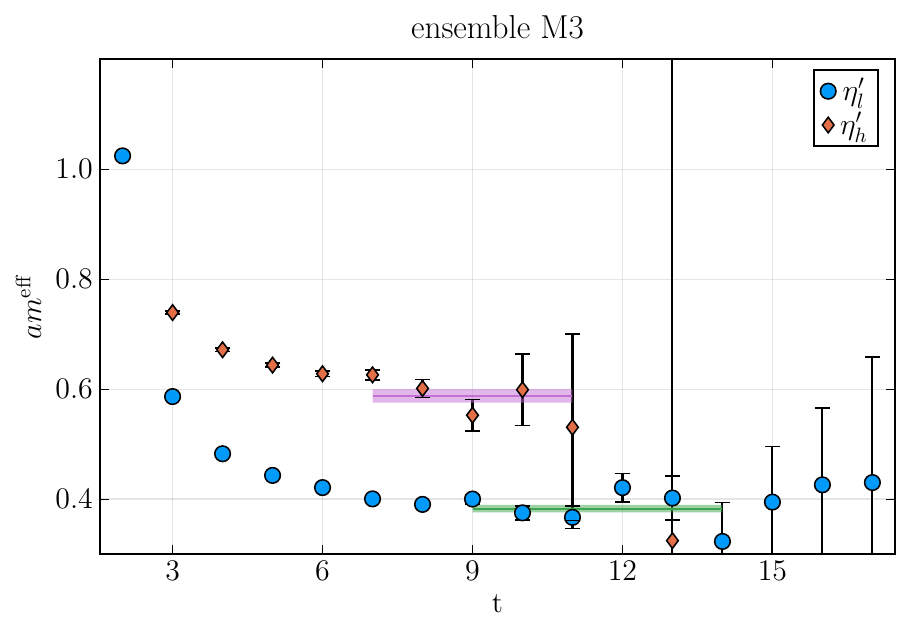}
    \includegraphics[width=.45\textwidth]{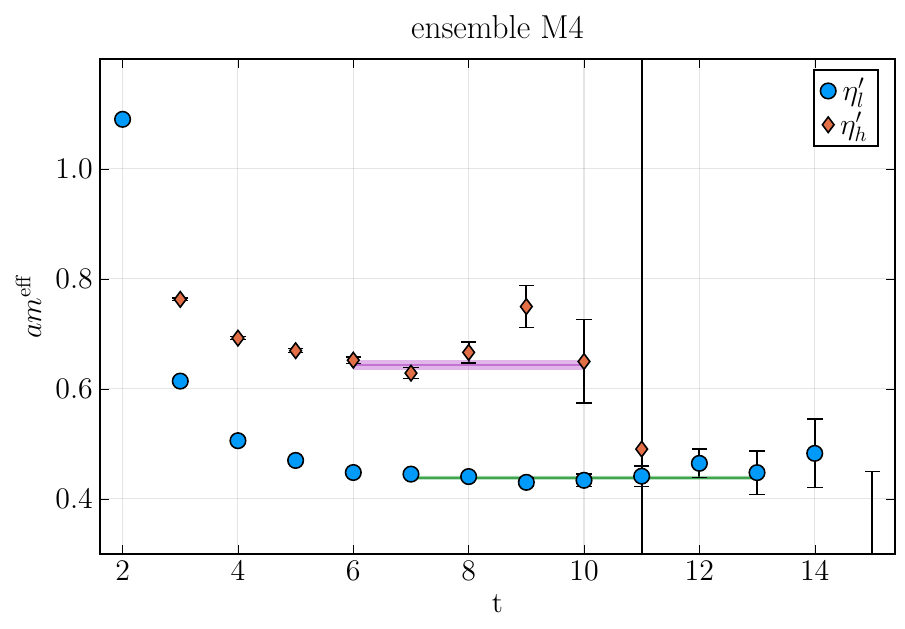}
    \includegraphics[width=.45\textwidth]{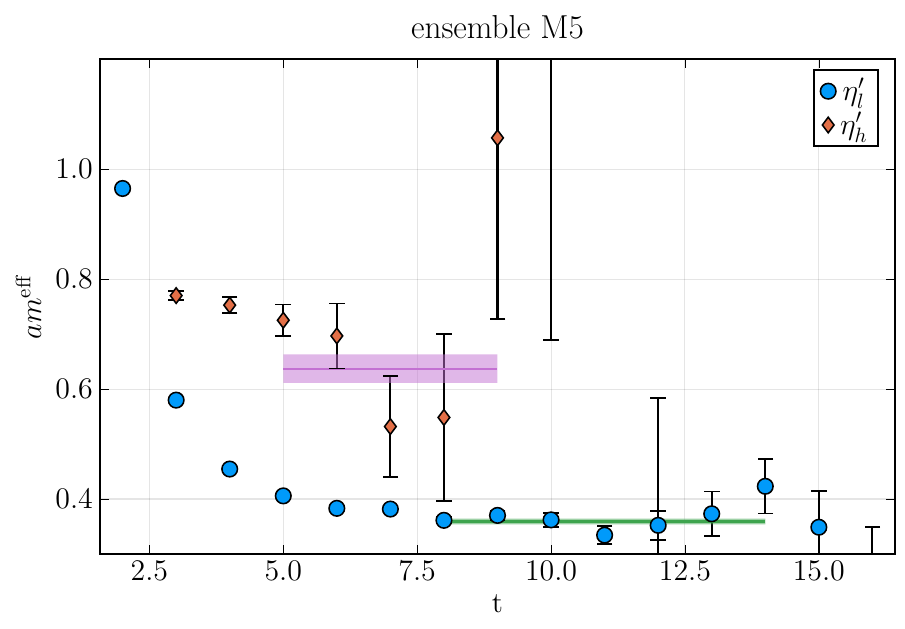}
    \caption{Effective masses, $a m^{\rm eff}_{\etalight}<a m^{\rm eff}_{\etaheavy}$ in units of the lattice spacing, of the ground state and first excited state in the flavor singlet, pseudoscalar meson sector of the $Sp(4)$ theory coupled to two Dirac fermions transforming in the fundamental representation, ${\rm f}$, and three in the antisymmetric, ${\rm as}$. The effective mass makes use of the eigenvalues extracted in the GEVP analysis. We display, for comparison, also the masses as extracted by an exponential fit to the eigenvalues as colored bands. The variational analysis used three distinct levels of smearing. For some ensembles, there is no clear plateau for the first excited state. The five panels correspond to the five available ensembles. }
    \label{fig:effective_masses}
\end{figure}

We present in this section our main numerical results. First, we explain our strategy in applying the GEVP analysis. As discussed in the previous section, we applied APE smearing to the ensembles,
as well as Wuppertal smearing to the meson operators. 
A basis of 18 distinct operators was obtained by varying the number smearing levels applied to the operators in Eqs.~\eqref{eq:ps_operator1} and \eqref{eq:ps_operator2}.
While the inclusion of further smearing levels in the GEVP improves the determination of the effective mass, it soon reintroduces a signal-to-noise ratio in the form of a loss signal at intermediate values of $t$. Hence, we
identify by inspection a subset of smearing levels appropriate to our analysis. 
By inspection, we find that the optimal choice for our purpose is to include only a subset of the operators produced with different $N^{\rm smear}$. 
We find, that restricting ourselves to three smearing levels $N^{\rm smear}=0,40,80$ is a good choice for our ensembles in all channels studied here. 
We observed that the stability of the GEVP analysis is improved by using a small value of $t_0=1$. This choice may in principle give rise to unaccounted systematic effects, that scale with $\mathcal O \left( e^{- \Delta E t_0} \right)$, where $\Delta E$ is the energy difference to the next state that is not properly captured by the chosen variational basis~\cite{Blossier:2009kd}. \rev{In appendix \ref{sec:t0_comparison} we show that no significant deviations are observed when choosing a larger value of $t_0$.} 

\subsection{Meson masses}
\label{Sec:results_mass}

In Fig.~\ref{fig:effective_masses}, we show the effective masses of the states, 
$a m^{\rm eff}_{\etalight}<a m^{\rm eff}_{\etaheavy}$, expressed in units of the lattice scale, $a$. The effective masses plots, obtained from the GEVP analysis, are shown to provide visual guidance in the choice of the fitting range. 
In some cases, only an approximate plateau appears within the statistical errors. This could be attributed to an insufficient stochastic sampling of the Dirac propagator. In this work we have chosen $n_{\rm src}=64$ stochastic sources to sample the disconnected diagrams in \eqref{eq:wick1}-\eqref{eq:wick3}.
This choice  introduces an error in the correlation matrix, that is expected to scale as
$\mathcal{O} \left( 1/\sqrt{n_{\rm src}} \right)$ \cite{Wilcox:1999ab}. 

As can be seen
in Fig.~\ref{fig:effective_masses}, the signal in the effective mass is lost for times $t$ much smaller than the temporal lattice midpoint $T/2$. Thus, we only fit a single exponential in the specified fit interval. The choices we made for the fitting parameters and the associated $\chi^2/N_{\rm d.o.f.}$ values are reported in Tab.~\ref{tab:fit_parameters}. The extracted meson masses (expressed in lattice units) are reported in Tab.~\ref{tab:fitresults}. We find that the statistical uncertainties obtained  for the flavor-singlet states are up to an order of magnitude larger than those for the flavored states. For the aforementioned reasons, it is likely that the systematic errors are also larger than those of the flavored mesons.

\begin{table}[t]
    \caption{Fit parameters used in the extraction of the meson masses, reported in Tab.~\ref{tab:fitresults}. We report the fitting ranges, $I$, the number of exponential terms, $N_{\rm exp}$, used in the fit, and the values of $\chi^2 / N_{\rm d.o.f.}$ for every fit.}
    \label{tab:fit_parameters}
    \begin{tabular}{|c|c|c|c|c|c|c|c|c|c|c|c|c|c|}
	\hline
   Label & $I_{\eta^{\prime}_l}$ & $I_{\eta^{\prime}_h}$ & $I_{\rm{PS}}$ & $I_{\rm{ps}}$ & $I_{\rm{V}}$ & $I_{\rm{v}}$ & $N_{\rm exp}$ & $\chi^2 / N_{\rm d.o.f.}$ & $\chi^2 / N_{\rm d.o.f.}$ & $\chi^2 / N_{\rm d.o.f.}$ & $\chi^2 / N_{\rm d.o.f.}$ & $\chi^2 / N_{\rm d.o.f.}$ & $\chi^2 / N_{\rm d.o.f.}$\\&&&&&&&&$ \eta^{\prime}_l$&$\eta^{\prime}_h$&${\rm PS}$&${\rm ps}$&$ {\rm V}$&$ {\rm v}$  \\
 \hline \hline 
	M1&(9,15)&(5,11)&(7,14)&(9,14)&(7,11)&(7,14)&1&2.7&2.4&3.8&2.9&0.9&1.5\\
	M2&(9,13)&(7,11)&(8,13)&(8,15)&(8,13)&(8,16)&1&1.8&1.9&2.7&1.7&2.3&1.8\\
	M3&(9,14)&(7,11)&(9,16)&(10,13)&(11,20)&(10,13)&1&1.4&1.4&1.1&1.9&1.3&1.8\\
	M4&(7,13)&(6,10)&(8,18)&(7,16)&(8,16)&(7,13)&1&2.2&2.0&1.3&2.0&1.3&3.4\\
	M5&(8,14)&(5,9)&(9,16)&(7,13)&(12,16)&(8,13)&1&1.4&1.6&1.4&2.2&1.5&1.9\\
	\hline\hline
\end{tabular}
\end{table}

\begin{table}[b]
    \caption{Meson masses extracted from large Euclidean-time behavior of the eigenvalues, $\lambda(t)$, within
     the GEVP analysis. We report the ground states and first excited states in the  flavor-singlet pseudoscalar
     channel as well as the ground states for flavored pseudoscalar and vector mesons made of either species of fermions, and for all five ensembles, which we characterize by the lattice coupling, $\beta$, its number of sites in the temporal, $N_t$, and spatial, $N_s$, directions, and the bare masses of the fermions in lattice units, $am_0^{\rm f}$ and $am_0^{\rm as}$.}
    \label{tab:fitresults}
    \begin{tabular}{|c|c|c|c|c|c|c|c|c|c|c|c|}
	\hline
	Label&$\beta$&$N_t$&$N_l$&$am_0^{\rm f}$&$am_0^{\rm as}$&$am_{\eta^{\prime}_l}$&$am_{\eta^{\prime}_h}$&$am_{\rm PS}$&$am_{\rm ps}$&$am_{\rm V}$&$am_{\rm v}$\\
	\hline\hline
	M1&6.5&48&20&-0.71&-1.01&0.3769(96)&0.6334(59)&0.3639(14)&0.6001(11)&0.4030(33)&0.6452(18)\\
	M2&6.5&64&20&-0.71&-1.01&0.3867(68)&0.619(13)&0.3648(13)&0.59856(82)&0.4038(17)&0.6421(15)\\
	M3&6.5&96&20&-0.71&-1.01&0.3826(67)&0.588(12)&0.3652(16)&0.59940(79)&0.4040(18)&0.6467(21)\\
	M4&6.5&64&20&-0.7&-1.01&0.4381(33)&0.6433(88)&0.4067(13)&0.62426(85)&0.4476(17)&0.6742(13)\\
	M5&6.5&64&32&-0.72&-1.01&0.3591(53)&0.637(26)&0.31076(68)&0.57718(85)&0.3518(12)&0.6223(15)\\
	\hline\hline
\end{tabular}
\end{table}

\begin{figure}
    \centering
    \includegraphics[width=.45\textwidth]{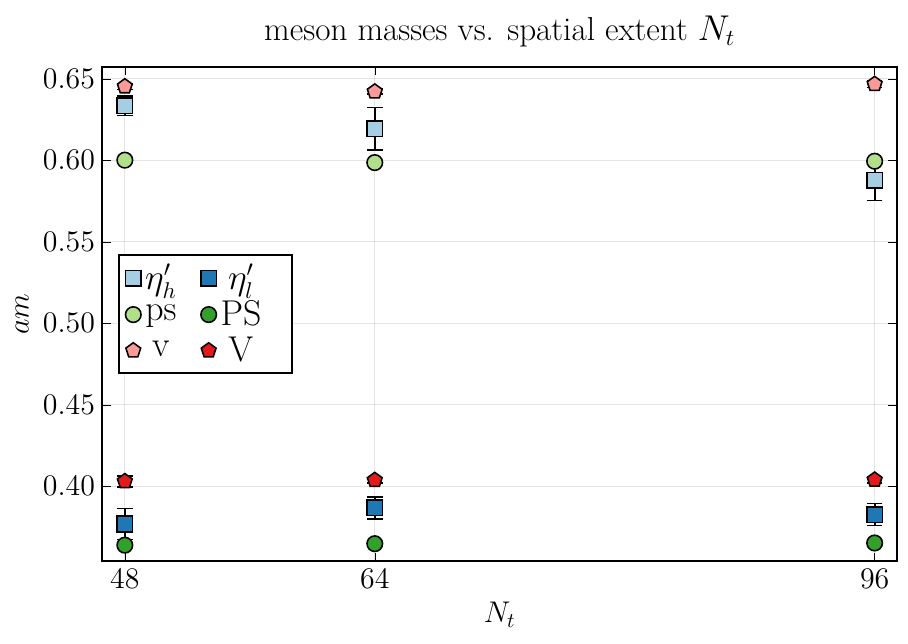}
    \includegraphics[width=.45\textwidth]{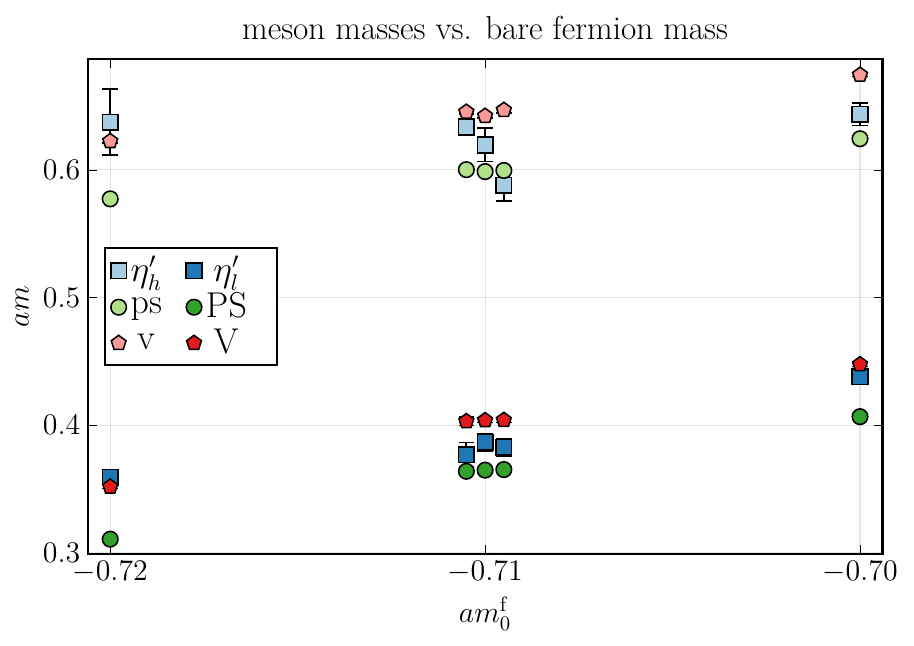}
    \includegraphics[width=.45\textwidth]{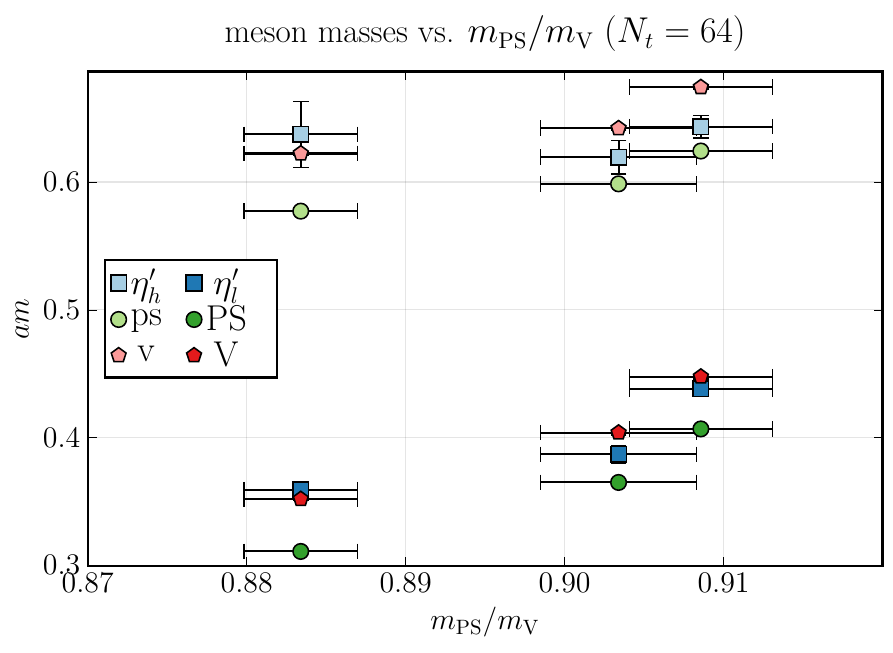}
    \includegraphics[width=.45\textwidth]{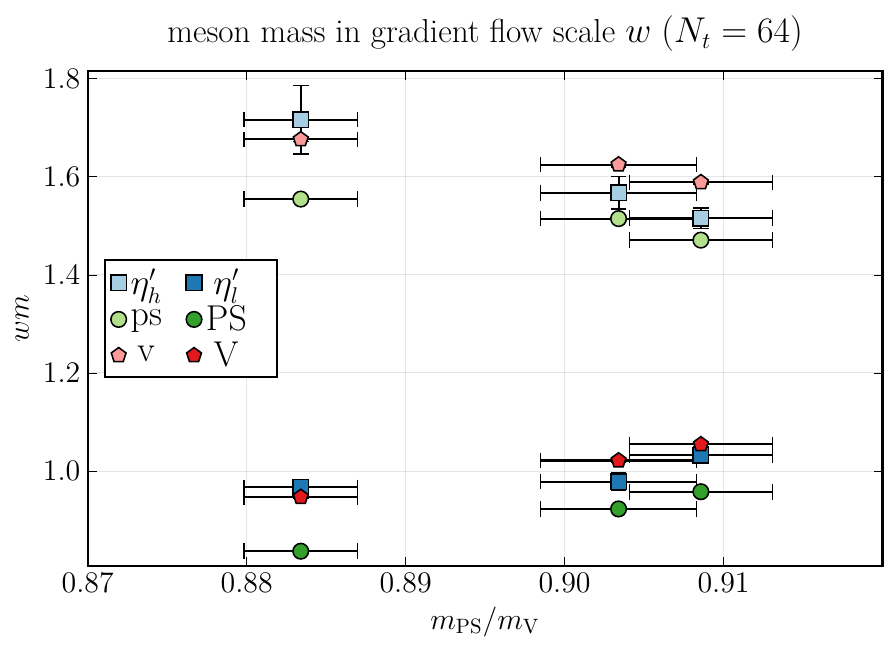}
    \caption{Masses, $a m$, of the lightest flavor-singlet and flavored pseudoscalar mesons, and flavored vector mesons, constituted of either species of fermions.  Top left panel: measurements of the masses in lattice units for the three ensembles (M1 - M3) that have common values of   $am_0^{\rm as}=-1.01$ and $am_0^{\rm f}=0.71$. The masses as shown for the three available choices of spatial lattice extent $N_t$. Top right panel: measurements of the  masses in lattice units for the three available choices of  bare fundamental fermion mass. A small horizontal offset has been introduced, for visual clarity, in the ensembles M1 - M3. Bottom left panel: measurement of the masses  in lattice units for all ensembles with $N_t=64$ (M2, M4, and M5),  plotted against the ratio of the mass of the flavored pseudoscalar and vector mesons constituted of fermions in the fundamental representation.
    Bottom right panel: same as bottom left, but the masses of the mesons are expressed in gradient flow units.
    \label{fig:spectrum}}
\end{figure}

In Fig.~\ref{fig:spectrum}, we show all the low-lying pseudoscalar and vector meson masses. They are displayed as a function of $N_t$ in the left-hand panel and as a function of $am_0^{\rm f}$ in the
right-hand panel. On the same plots, we display the masses of the corresponding states in the flavored channels. Overall, the ground states and first excited state masses in the flavor-singlet channels are found to lie, within uncertainties, in the range determined by the masses corresponding flavored channels.
We find larger uncertainties for the $\eta'$ state, hence the mass hierarchy is not fully determined. For the lightest ensemble, M5, the flavor-singlet pseudoscalar states have mass compatible to the flavored vector mesons.

The general trends exhibited by our results suggest that the effect of the disconnected diagrams is suppressed by the heavy fermion masses present in these ensembles. Thus, the spectrum resembles that of $N_{\rm  f}=2$ single-representation theories at moderately-heavy to heavy fermion mass~\cite{Bennett:2023rsl}. A suppression of the disconnected diagrams also implies a suppression of the $\etalight-\etaheavy$ mixing effects, according to Eq.~\eqref{eq:wick3}. Thus, we expect that the associated mixing angle is small and that the state $a$ is mostly dominated by the fundamental fermionic contribution, whereas the mass of the $\eta^{\prime}$ is mostly determined by the antisymmetric fermion masses. We will return to the determination of the mixing angle in the next subsection.

Because the ensembles M1-M3 share the same lattice parameters, with only the temporal lattice 
extent, $N_t$, distinguishing them, we can compare across the three to ascertain the contribution of $N_t$ to systematic effects. We find that for both the singlet ground states, $\etalight$, and the first excited state, $\etaheavy$, measurements with different $N_t$ yield compatible results. This may not be the case for further excited states~\cite{Bennett:2024cqv}, but these considerations go beyond the purposes of the present study.

\subsection{Mixing angle}
\label{Sec:results_angle}

\begin{figure}
    \centering
    \includegraphics[width=.45\textwidth]{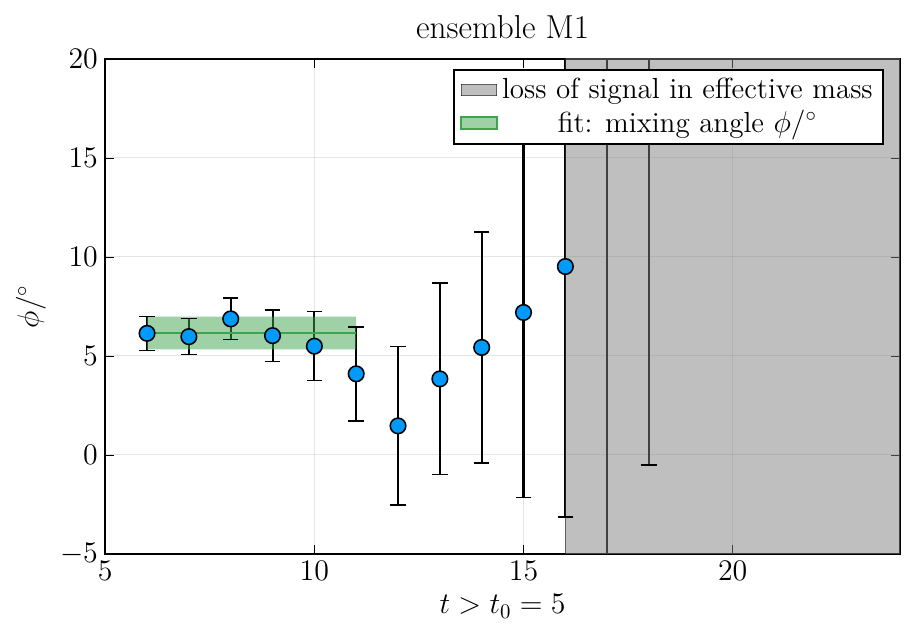}
    \includegraphics[width=.45\textwidth]{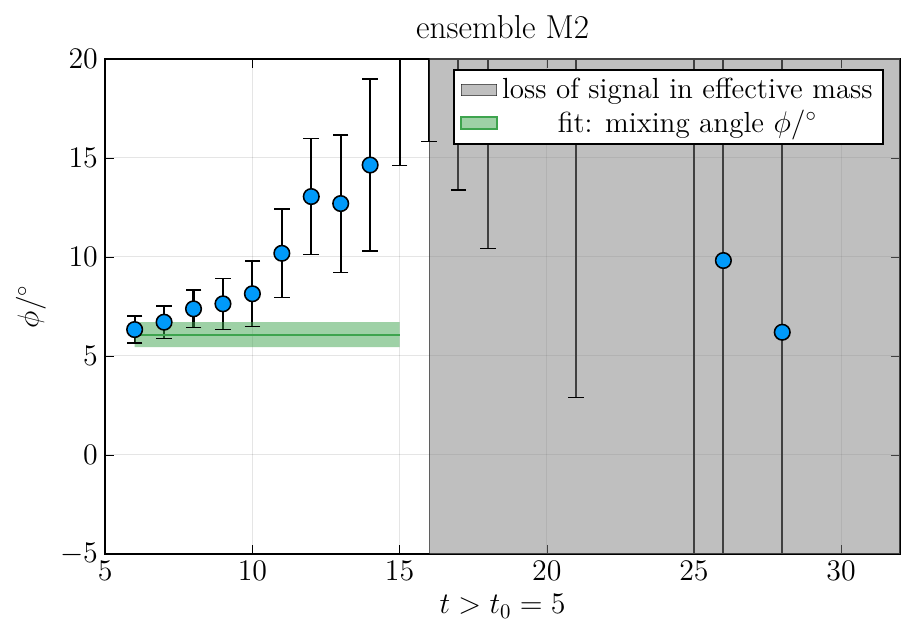}
    \includegraphics[width=.45\textwidth]{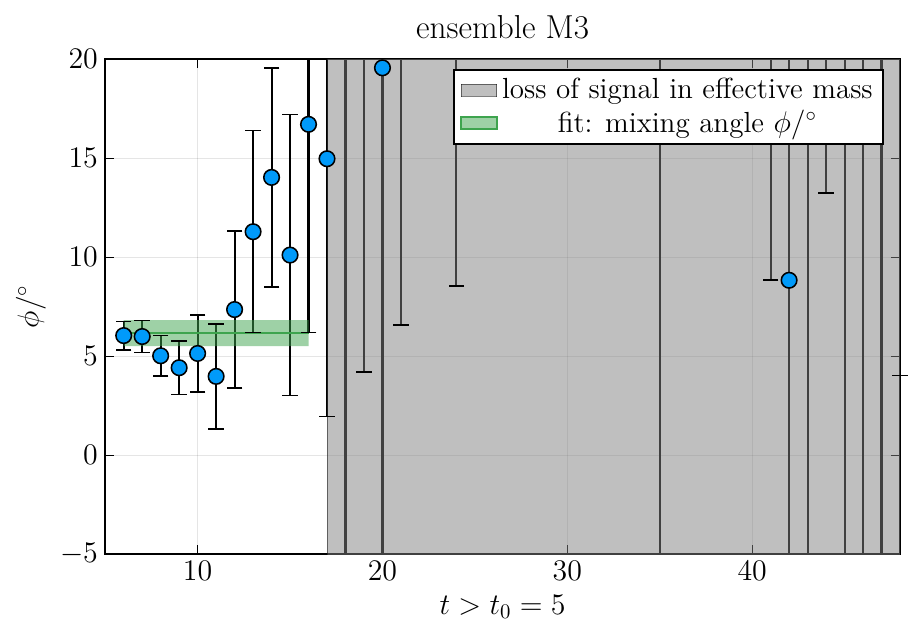}
    \includegraphics[width=.45\textwidth]{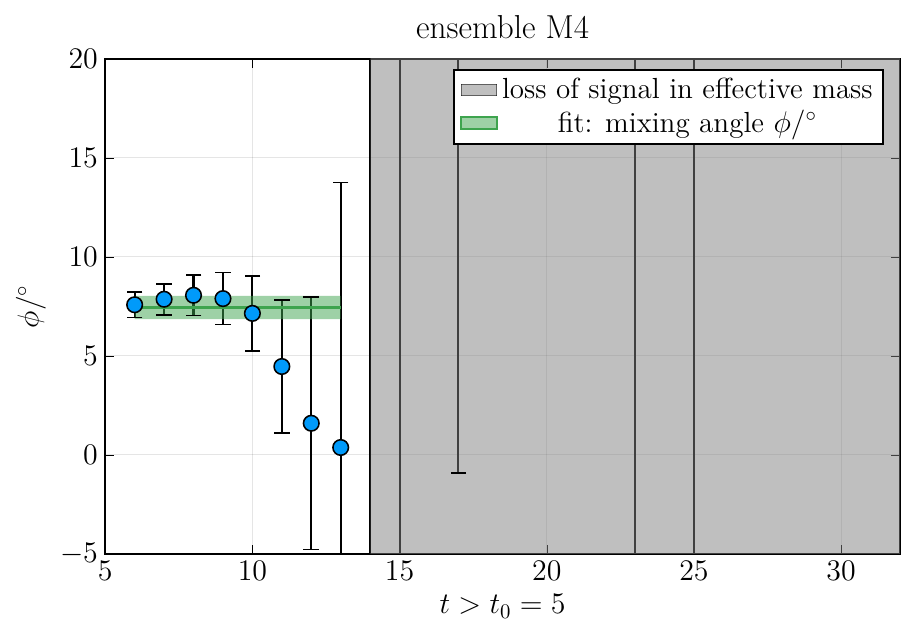}
    \includegraphics[width=.45\textwidth]{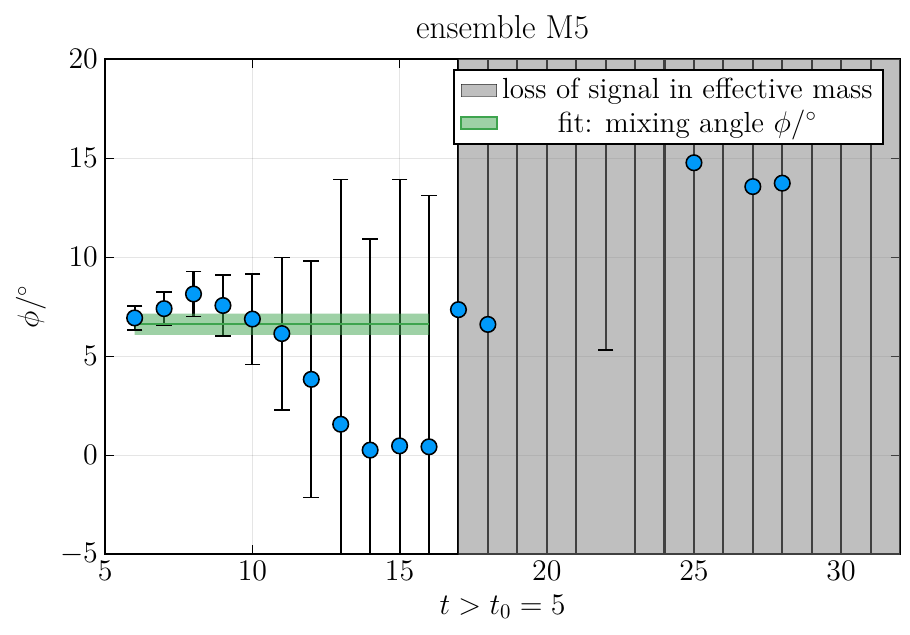}
    \caption{Effective mixing angles, $\phi$,  between the lightest pseudo-scalar singlets,
     for all ensembles M1-M5, based on the pseudoscalar singlet matrix elements defined in Eq.~\eqref{eq:PS_matrix_elements}. We fit the effective mixing angle in the range from a minimum value $t=t_0+1$ until a maximum value  $t=t_{\rm max}$. Conventionally, we choose the latter  to coincide with  the smallest $t$ value at which the relative error of the ground-state effective mass is larger than 50\%. 
     We find good signal-to-noise ratios in all ensembles, even for the comparatively large value of $t_0 =5$. 
     We find no clear evidence of mass dependence of the mixing angle.\label{fig:effective_mixing_angle}}
\end{figure}

With the available ensembles we measured the pseudoscalar matrix elements of Eq.~\eqref{eq:PS_matrix_elements}, from which we extracted the effective mixing angle, $\phi$,  determined according to Eq.~\eqref{eq:effective_mixing_angle}, which we display in Fig.~\ref{fig:effective_mixing_angle} as a function of the time $t$. A similar analysis for the axial-vector currents, in Eq.~\eqref{eq:matrix_elements}, did not yield a significant signal, hence we did not determine the pseudoscalar decay constants. This behavior is similar from what has been found in QCD, where the flavor-singlet axial-vector  matrix elements are affected by poor signal-to-noise ratios~\cite{Ottnad:2017bjt}. 

We stress that the result for the effective mixing angle displayed in Fig.~\ref{fig:effective_mixing_angle} have been obtained with a simplified analysis that only involves local sources. The simpler GEVP is analyzed with  a larger value of $t_0=5$, as it is in this regime where the systematic error on the determination of the matrix element should be smaller, see Ref.~\cite{Blossier:2009kd}. We leave to future, high-precision studies the task of measuring  the relevant components of the correlation matrix itself~\cite{Bali:2021qem} to determine simultaneously the mixing angle and the decay constants.

We extract the mixing angle, $\phi$, via a constant fit to the effective mixing angle. \rev{We similarly determine $\phi_\etalight$ and $\phi_\etaheavy$.} We choose the fitting interval to start at $t=t_0+1$ and end at the first $t=t_{\rm max}$ for which the relative uncertainty in the effective mass of the flavor-singlet pseudoscalar ground state exceeds $50 \%$. We report the extracted mixing angles in Tab.~\ref{tab:mixing_results}. We find a small mixing angle in all available ensembles. \rev{We find that a parametrization using two mixing angles leads to statistically significant deviations.}  We do not find clear evidence of  mass dependence in the mixing angles \rev{$\phi$ and $\phi_\etalight$} in the ensembles studied here. All measurements indicate a mixing angle of roughly $6^{\circ}$ for $\phi$. These results further support earlier conclusions: because of the large value of the fermion masses, the mixing effects between the flavor-singlet pseudoscalar states $\etalight$ and $\etaheavy$ are strongly suppressed. 

\begin{table}[t]
    \caption{Results for the pseudoscalar singlet mixing angle, $\phi$, expressed in degrees,  extracted from the flavor-singlet pseudoscalar matrix elements, using correlation functions with point-like source and sink (no Wuppertal smearing). The  quoted uncertainties have been obtained using jackknife resampling.} 
    \label{tab:mixing_results}
    \rev{\begin{tabular}{|c|c|c|c|c|c|c|}
	\hline
    Label&$~~~~\beta~~~~$&$~~~~N_t~~~~$&$~~~~N_s~~~~$&$~~~~\phi/{}^{\circ}~~~~$&$~~~~\phi_\etalight/{}^{\circ}~~~~$&$~~~~\phi_\etaheavy/{}^{\circ}~~~~$ \\ \hline \hline 
	M1&6.5&48&20&6.15(83)&3.83(57)&9.8(1.1)\\
	M2&6.5&64&20&6.07(63)&3.74(43)&9.78(89)\\
	M3&6.5&96&20&6.16(66)&3.76(44)&10.00(92)\\
	M4&6.5&64&20&7.44(58)&4.77(42)&12.26(86)\\
	M5&6.5&64&32&6.61(54)&5.87(52)&7.67(64)\\
	\hline\hline
\end{tabular}}
\end{table}

\section{Summary and Outlook}
\label{Sec:S}

In this paper, we have presented the first numerical lattice study of the mass spectrum of flavor-singlet pseudoscalar
bound states in an $Sp(4)$ gauge theory with fermions in multiple representations. The channels of interest result from the mixing  effects between the Abelian PNGBs associated with the two global $U(1)$ factors in the approximate symmetry of the system. These Abelian symmetries are broken explicitly both by the masses of the fermions, and by the axial anomaly. The motivation for this study is that, in the context of composite Higgs models, the phenomenology of such flavor singlets deserves special consideration~\cite{Belyaev:2016ftv}, and requires detailed knowledge of non-perturbative properties of the theory in its strongly coupled regime.
 
Our measurements have been performed on five available ensembles, generated in a lattice field theory that is of interest in its own terms, as it has been proposed as a short-distance completion to the minimal CHM that also implements top partial compositeness~\cite{Barnard:2013zea}.

This type of spectroscopy measurement is  challenging because it requires performing explicit calculations of disconnected diagrams contributing to the two-point functions of interest, which introduce high noise level in the numerical analysis.
These measurements were obtained with a combination of APE and Wuppertal smearing, as well as with an implementation of the GEVP analysis. 

The results presented above provide the first determination--and a demonstration of feasibility--of the masses of the ground and first excited state in the flavor-singlet
pseudoscalar channel of this theory, as well as of their associated mixing angle.
 
Some limitations of the analysis descend from the fact that all the ensembles have the same value of the lattice coupling, $\beta$, so that a continuum-limit extrapolation is beyond the reach of this analysis.
Furthermore,  the  masses for the two species of fermions belong to a regime in which 
the theory is far from the massless limit.
As a result, the masses of neither the flavor-singlet nor the flavored pseudoscalar states in the theory are particularly light,
and a massless extrapolation of the results is also beyond current reach.

 The natural next step for future studies would be to deploy the numerical strategy we developed and 
 tested in a large-scale 
 investigation that would allow performing continuum and massless extrapolations.
 This endeavor would allow to better assess the effect on the spectrum and mixing angle of the axial anomaly, 
 and connect this study to ongoing model-building and collider phenomenology programs.
 Furthermore, having demonstrated the viability of the numerical techniques, it would be interesting to understand 
 how the results depend also on other intrinsic parameters of the theory, such as the 
 numbers, $N_{\rm f}$ and $N_{\rm as}$, of fermions transforming on the fundamental and antisymmetric representation of the group. An interesting connection to more formal field-theory research 
 topics would involve  changing the group, within the $Sp(2N)$ class of theories,
to explore the approach to the large-$N$ limit.
We plan to exploit all  of these opportunities for further research in the future.

\begin{acknowledgments}

We would like to thank Giacomo Cacciapaglia, Gabriele Ferretti, Thomas Flacke, Anna Hasenfratz, Chulwoo Jung, and Sarada Rajeev, for useful discussions during the "PNU Workshop on Composite Higgs: Lattice study and all”, at Haeundae, Busan, in February 2024, where preliminary results of this study were presented.

The work of EB and BL is supported in part by the EPSRC ExCALIBUR programme ExaTEPP (project EP/X017168/1). The work of EB, BL, MP, and FZ has been supported by the STFC Consolidated Grant No. ST/X000648.
The work of EB has also been supported by the UKRI Science and Technology Facilities Council (STFC) Research Software Engineering Fellowship EP/V052489/1.
The work of NF has been supported by the STFC Consolidated Grant No. ST/X508834/1.
The work of DKH was supported by Basic Science Research Program through the National Research Foundation of Korea (NRF) funded by the Ministry of Education (NRF-2017R1D1A1B06033701). The work of DKH was further supported by the National Research Foundation of Korea (NRF) grant funded by the Korea government (MSIT) (2021R1A4A5031460).
The work of JWL is supported by IBS under the project code, IBS-R018-D1. 
The work of HH and CJDL is supported by the Taiwanese MoST grant 109-2112-M-009-006-MY3 and NSTC grant 112-2112-M-A49-021-MY3. The work of CJDL is also supported by Grants No. 112-2639-M-002-006-ASP and No. 113-2119-M-007-013-.
The work of BL and MP has been further supported in part by the STFC  Consolidated Grant No. ST/T000813/1.
BL and MP received funding from the European Research Council (ERC) under the European Union’s Horizon 2020 research and innovation program under Grant Agreement No.~813942. 
The work of DV is supported by STFC under Consolidated Grant No. ST/X000680/1.

Numerical simulations have been performed on the DiRAC Extreme Scaling service at the University of Edinburgh, and on the DiRAC Data Intensive service at Leicester.
The DiRAC Extreme Scaling service is operated by the Edinburgh Parallel Computing Centre on behalf of the STFC DiRAC HPC Facility (www.dirac.ac.uk). This equipment was funded by BEIS capital funding via STFC capital grant ST/R00238X/1 and STFC DiRAC Operations grant ST/R001006/1. DiRAC is part of the National e-Infrastructure

{\bf Research Data Access Statement}---The data generated for this manuscript can be downloaded from  Ref.~\cite{data_release} and the analysis code from Ref~\cite{analysis_release}. 

{\bf Open Access Statement}---For the purpose of open access, the authors have applied a Creative Commons  Attribution (CC BY) licence  to any Author Accepted Manuscript version arising.

\end{acknowledgments}

\appendix

\section{Systematic effects related to the choice \texorpdfstring{$t_0$}{t0}}\label{sec:t0_comparison}

\rev{
In principle, the value of $t_0$ in the GEVP should be chosen to be large in order to suppress systematic effects \cite{Blossier:2009kd}. However, a choice of small $t_0$ can lead to smaller statistical uncertainties. In Tab.~\ref{tab:t0_comparison} we report the masses of the pseudoscalar singlet mesons as a function of $t_0$. We do not find any significant deviations.} 
\begin{table}[t]
    \caption{\rev{The mass of the pseudoscalar singlet states as a function of $t_0$. When varying $t_0$ we find no deviations outside of statistical uncertainties.}} 
    \label{tab:t0_comparison}
    \rev{\begin{tabular}{|c|c|c|c|c|c|c|c|c|c|c|}
	\hline
   Label & $\beta$ & $N_t$ & $N_l$ & $am_0^{\rm f}$ & $am_0^{\rm as}$ & $am_{\eta^{\prime}_h} (t_0=1)$ & $am_{\eta^{\prime}_h} (t_0=2)$ & $am_{\eta^{\prime}_h} (t_0=3)$ & $am_{\eta^{\prime}_h} (t_0=4)$ & $am_{\eta^{\prime}_h} (t_0=5)$   \\
 \hline \hline 
	M1&6.5&48&20&-0.71&-1.01&0.6354(61)&0.6344(62)&0.6306(74)&0.6306(77)&0.634(10)\\
	M2&6.5&64&20&-0.71&-1.01&0.621(13)&0.6239(74)&0.6225(92)&0.623(10)&0.623(11)\\
	M3&6.5&96&20&-0.71&-1.01&0.592(13)&0.6111(76)&0.6079(92)&0.609(10)&0.620(18)\\
	M4&6.5&64&20&-0.7&-1.01&0.6436(89)&0.6423(90)&0.643(11)&0.647(12)&0.658(13)\\
	M5&6.5&64&32&-0.72&-1.01&0.667(26)&0.660(28)&0.668(21)&0.659(21)&0.651(20)\\
	\hline\hline
   Label & $\beta$ & $N_t$ & $N_l$ & $am_0^{\rm f}$ & $am_0^{\rm as}$ & $am_{\eta^{\prime}_l} (t_0=1)$ & $am_{\eta^{\prime}_l} (t_0=2)$ & $am_{\eta^{\prime}_l} (t_0=3)$ & $am_{\eta^{\prime}_l} (t_0=4)$ & $am_{\eta^{\prime}_l} (t_0=5)$   \\
 \hline \hline 
	M1&6.5&48&20&-0.71&-1.01&0.3769(96)&0.3968(34)&0.3900(43)&0.3836(68)&0.3913(69)\\
	M2&6.5&64&20&-0.71&-1.01&0.3867(68)&0.3969(34)&0.3917(43)&0.3898(51)&0.3912(70)\\
	M3&6.5&96&20&-0.71&-1.01&0.3826(67)&0.3876(34)&0.3826(41)&0.3815(48)&0.3750(84)\\
	M4&6.5&64&20&-0.7&-1.01&0.4380(33)&0.4394(26)&0.4362(29)&0.4348(45)&0.4360(36)\\
	M5&6.5&64&32&-0.72&-1.01&0.3590(53)&0.3582(54)&0.3617(33)&0.3586(63)&0.3615(36)\\
	\hline\hline
\end{tabular}}
\end{table}

\bibliographystyle{JHEP} 
\bibliography{ref}
 
\end{document}